\documentclass[epj,nopacs]{svjour}
\usepackage{epsfig,array}
\usepackage{graphicx,widetext}
\def\nn{\nonumber \\}

\def\fun#1#2{\lower3.6pt\vbox{\baselineskip0pt\lineskip.9pt
\ialign{$\mathsurround=0pt#1\hfil##\hfil$\crcr#2\crcr\sim\crcr}}}

\begin{document}
\newcommand{\be}{\begin{eqnarray}}
\newcommand{\ee}{\end{eqnarray}}
\newcommand{\beq}{\begin{equation}}
\newcommand{\eeq}{\end{equation}}
\newcommand{\inli}{\int\limits}


\date{10.10.2020}

\title{\bf
Partial decay widths of baryons in the spin-momentum operator expansion method}

\author
{A.V. Sarantsev \inst{1,2}}

\institute {$^1$ HISKP, Bonn University, 53115 Bonn, Germany\\
$^2$ National Research Center "Kurchatov Institute", PNPI, Gatchina,
188300, Russia}


\abstract{The covariant operator expansion method used by the
Bonn-Gatchina group for the analysis of the meson photoproduction
data is extended on the case of meson electro-production reactions.
The angular dependence of the partial waves is deduced and the
obtained amplitudes are compared with those used in other analyses
of the electro-production reactions.}



\mail{andsar@hiskp.uni-bonn.de}

\maketitle

\section{Introduction}

Reactions with pseudoscalar mesons in the final state provides the
main part of the information about spectrum and properties of hadron
resonances.  The final states with pseudoscalar mesons only are easy
to measure  and the data are relatively easy to analyze: for
example, in the case of pion-pion scattering only measurement of the
differential cross section provides the full information about
partial wave amplitudes. In many cases such analysis can be
performed in two steps. At the first step the measured angular
distribution is analyzed at fixed energy and partial wave amplitudes
are extracted with some precision. At second step the energy
dependence of the extracted amplitudes is analyzed and their
analytical structure is determined.

However, in the case of photoproduction of the mesons off nucleon
the analysis of the data is a more complicated issue. To perform the
full analysis of a single pseudoscalar meson production reaction at
least eight independent polarization observables should be measured.
In many cases such information is not available and the data are
analyzed in so called energy dependent approach. Here the partial
waves are extracted from the simultaneous analysis of the energy and
angular distributions. Thus an observation of a resonance in a
particular partial wave notably reduces the number of parameters and
the resonance properties can be defined from the restricted number
of measured observables. Moreover the great advantage of such
approach is the possibility for a combined analysis of reactions
with different final (or initial) states. In this case the
polarization observables measured in one reaction can provide a key
information for the analysis of the data measured in another
reaction.

Another important source for the information about baryon resonances
and their properties is the meson electro-production reactions. Such
data allows us to study the dependence of the resonance production
couplings on the mass of the virtual photon and therefore about size
and internal structure of the resonances. It is a vital information
which can help to understand the nature of the baryon states and
properties of the strong interactions.

Originally the fully covariant Bonn-Gatchina formalism was developed
for the analysis of the meson photoproduction reactions: it is
described in details in the paper \cite{Anisovich:2007zz}. This
approach was successfully used for the analysis of the data measured
by the CB-ELSA, CLAS and MAMI collaborations. It also was applied by
the HADES collaboration for the analysis of the pion induced meson
production data. In this paper the Bonn-Gatchina approach is
extended for the analysis of the electro-production data. Our
formalism is compared with the covariant approach
\cite{Krivoruchenko:2001jk} suggested earlier.

\section{Decay of the resonance into two spinless particles}

The orbital angular momentum operators for $L \le 3 $ are:
\begin{eqnarray}
X^{(0)}&=&1\ , \qquad X^{(1)}_\mu=k^\perp_\mu\ , \qquad\nonumber \\
X^{(2)}_{\mu_1 \mu_2}&=&\frac32\left(k^\perp_{\mu_1}
k^\perp_{\mu_2}-\frac13\, k^2_\perp g^\perp_{\mu_1\mu_2}\right), \nonumber  \\
X^{(3)}_{\mu_1\mu_2\mu_3}&=&\frac52\Big[k^\perp_{\mu_1} k^\perp_{\mu_2 }
k^\perp_{\mu_3}
\nn
&-&\frac{k^2_\perp}5\left(g^\perp_{\mu_1\mu_2}k^\perp
_{\mu_3}+g^\perp_{\mu_1\mu_3}k^\perp_{\mu_2}+
g^\perp_{\mu_2\mu_3}k^\perp_{\mu_1}
\right)\Big]\,.~~~
\label{x-oper}
\ee
The operators $X^{(L)}_{\mu_1\ldots\mu_L}$ for $L\ge 1$ can be
written in the form of the recurrence expression:
\be
X^{(L)}_{\mu_1\ldots\mu_L}&=&k^\perp_\alpha
Z^{\alpha}_{\mu_1\ldots\mu_L} \; ,
\nonumber\\
Z^{\alpha}_{\mu_1\ldots\mu_L}&=&
\frac{2L-1}{L^2}\Big (
\sum^L_{i=1}X^{{(L-1)}}_{\mu_1\ldots\mu_{i-1}\mu_{i+1}\ldots\mu_L}
g^\perp_{\mu_i\alpha}-
\nonumber \\
 \frac{2}{2L-1}  \sum^L_{i,j=1 \atop i<j}
&g^\perp_{\mu_i\mu_j}&
X^{{(L-1)}}_{\mu_1\ldots\mu_{i-1}\mu_{i+1}\ldots\mu_{j-1}\mu_{j+1}
\ldots\mu_L\alpha} \Big )\,.
\label{z}
\ee
Other useful properties of the orbital momentum operators are listed in
Appendix.

The projection operator $O^{\mu_1\ldots\mu_L}_{\nu_1\ldots \nu_L}$
is constructed from the metric tensors $g^\perp_{\mu\nu}$ and
has the following properties:
\be
X^{(L)}_{\mu_1\ldots\mu_L}
O^{\mu_1\ldots\mu_L}_{\nu_1\ldots \nu_L}\
&=&\ X^{(L)}_{\nu_1\ldots \nu_L}\ , \nonumber \\
O^{\mu_1\ldots\mu_L}_{\alpha_1\ldots\alpha_L} \
O^{\alpha_1\ldots\alpha_L}_{\nu_1\ldots \nu_L}\
&=& O^{\mu_1\ldots\mu_L}_{\nu_1\ldots \nu_L}\ .
\label{proj_op}
\ee
The projection operator projects any tensor with $n$ indices onto
tensors which satisfy the properties (\ref{x-oper}). For the lowest
states,
\be
O\!&=&\! 1\qquad O^\mu_\nu\!=\!g_{\mu\nu}^\perp
\nn
O^{\mu_1\mu_2}_{\nu_1\nu_2}\!&=&\!
\frac 12 \left (
g_{\mu_1\nu_1}^\perp  g_{\mu_2\nu_2}^\perp \!+\!
g_{\mu_1\nu_2}^\perp  g_{\mu_2\nu_1}^\perp  \!- \!\frac 23
g_{\mu_1\mu_2}^\perp  g_{\nu_1\nu_2}^\perp \right )\,.~
\ee
For higher states, the operator can be calculated using the
recurrent expression:
\be &&O^{\mu_1\ldots\mu_L}_{\nu_1\ldots
\nu_L}=\frac{1}{L^2} \bigg (
\sum\limits_{i,j=1}^{L}g^\perp_{\mu_i\nu_j}
O^{\mu_1\ldots\mu_{i-1}\mu_{i+1}\ldots\mu_L}_{\nu_1\ldots
\nu_{j-1}\nu_{j+1}\ldots\nu_L}-
\nonumber \\
 &&  \frac{4}{(2L-1)(2L-3)} \times
\nn    &&
\sum\limits_{i<j\atop k<m}^{L}
g^\perp_{\mu_i\mu_j}g^\perp_{\nu_k\nu_m}
O^{\mu_1\ldots\mu_{i-1}\mu_{i+1}\ldots\mu_{j-1}\mu_{j+1}\ldots\mu_L}_
{\nu_1\ldots\nu_{k-1}\nu_{k+1}\ldots\nu_{m-1}\nu_{m+1}\ldots\nu_L}
\bigg )\,.
\ee
The tensor part of the boson propagator is defined by the projection
operator. Let us write it as
\be
F^{\mu_1\ldots\mu_L}_{\nu_1\ldots\nu_L}=
(-1)^L\,O^{\mu_1\ldots\mu_L}_{\nu_1\ldots \nu_L}\,.
\label{boson_prop}
\ee


\section{The structure of the fermion propagator}

The wave function of a fermion is described as Dirac $\;$ bispinor,
as object in Dirac space represented by $\gamma$ matrices. In the
standard representation the $\gamma$ matrices have the following
form:
\be
\gamma_0=\left ( \begin{array}{cc} 1 & 0 \\ 0 & -1 \end{array}
\right ),\qquad \vec \gamma=\left (
\begin{array}{cc}
0 & \vec \sigma \\
-\vec \sigma & 0
\end{array}
\right ),\qquad \gamma_5=\left (
\begin{array}{cc}
0 & 1 \\
1 & 0
\end{array}
\right )\qquad
\ee
where $\vec \sigma$ are $2\times 2$ Pauli matrices. In this
representation the spinors for fermion particles with momentum p
are:
\be
u(p)=\frac{1}{\sqrt{p_0+m}} \left (
\begin{array}{c}
(p_0+m)\omega \\
(\vec p\vec \sigma)\omega
\end{array}
\right ), \nonumber \\
\bar u(p)=\frac{ \left ( \omega^* (p_0+m), -\omega^*(\vec p\vec
\sigma) \right ) }{\sqrt{p_0+m}}.
\ee
Here $\omega$ represents a 2-dimensional spinor and $\omega^*$ the
conjugated and transposed spinor. The normalization condition can be
written as:
\be \bar u(p) u(p)=2m  \quad
\sum\limits_{polarizations} u(p) \bar u(p)=m+\hat p
\label{bisp_norm}
\ee
We define $\hat p =p^\mu\gamma_\mu$.

The structure of the fermion propagator
$\mathcal P^{\mu_1\ldots\mu_n}_{\nu_1\ldots\nu_n}$
 was considered in details in
\cite{Anisovich:2007zz}. The propagator is defined as
\be
\mathcal P^{\mu_1\ldots\mu_n}_{\nu_1\ldots\nu_n}\!=
\frac{F^{\mu_1\ldots\mu_n}_{\nu_1\ldots\nu_n}}{M^2 -s -iM\Gamma}\,,
\ee
where
\be
F^{\mu_1\ldots\mu_n}_{\nu_1\ldots\nu_n}\!=\!(-1)^n
\frac{\sqrt{s}\!+\!\hat P}{2\sqrt{s}}
O^{\mu_1\ldots\mu_n}_{\xi_1\ldots \xi_n}
T^{\xi_1\ldots\xi_n}_{\beta_1\ldots \beta_n}
O^{\beta_1\ldots \beta_n}_{\nu_1\ldots\nu_n}\,.
\label{fp}
\ee
Here, $(\sqrt s+\hat P)$ corresponds to the numerator of fermion
propagator describing the particle with $J=1/2$ and
$n\!=\!J\!-\!1/2$ ($\sqrt s\!=\!M$ for the stable particle). We
define
\be T^{\xi_1\ldots\xi_n}_{\beta_1\ldots \beta_n}&=&
\frac{n+1}{2n\!+\!1} \big( g_{\xi_1\beta_1}\!-\!
\frac{n}{n\!+\!1}\sigma_{\xi_1\beta_1} \big)
\prod\limits_{i=2}^{n}g_{\xi_i\beta_i}, \nn
\sigma_{\alpha_i\alpha_j}&=&\frac 12
(\gamma_{\alpha_i}\gamma_{\alpha_j}-
\gamma_{\alpha_j}\gamma_{\alpha_i}).
\label{t1}
\ee
As in \cite{Anisovich:2007zz}, we introduced the factor $1/(2\sqrt
s)$ in the propagator which removes the divergency of this function
at large energies. For the stable particle it means that bispinors
are normalized as follows:
\be \bar u(k_N) u(k_N)\!=\!1\;,\;\;
\sum\limits_{polarizations}\!\!\!\!\!\! u(k_N)\bar u(k_N)
\!=\!\frac{m\!+\!\hat k_N}{2m}\;.
\label{bisp_pr}
\ee
Here and below, $\hat k\equiv\gamma_\mu k_\mu$.

It is useful to list the properties of the fermion propagator:
\be
&&P_{\mu_i}F^{\mu_1\ldots\mu_n}_{\nu_1\ldots \nu_n}
=P_{\nu_j}F^{\mu_1\ldots\mu_n}_{\nu_1\ldots \nu_n}=0\;,
\nn
&&\gamma_{\mu_i}F^{\mu_1\ldots\mu_n}_{\nu_1\ldots \nu_n}
=F^{\mu_1\ldots\mu_n}_{\nu_1\ldots \nu_n}\gamma_{\nu_j}=0\;,
\nn
&&F^{\mu_1\ldots\mu_n}_{\alpha_1\ldots \alpha_n}
F^{\alpha_1\ldots \alpha_n}_{\nu_1\ldots \nu_n}=
(-1)^n F^{\mu_1\ldots\mu_n}_{\nu_1\ldots \nu_n} \;,
\nn
&&\hat P F^{\mu_1\ldots\mu_n}_{\nu_1\ldots \nu_n}
=\sqrt s F^{\mu_1\ldots\mu_n}_{\nu_1\ldots \nu_n}.
\label{F_proper}
\ee

\subsection{\boldmath $\pi N$ vertices}

The states with $J=L\!+\!1/2$, where $L$ is the orbital momentum of the
$\pi N$ system, are called '+' states
($1/2^-$, $3/2^+$, $5/2^-$,\ldots).
The states with $J=L\!-\!1/2$ are called '-' states
($1/2^+$, $3/2^-$, $5/2^+$,\ldots).
The correspondent vertices are ($n=J\!-\!1/2$):
\be
N^+_{\mu_1\ldots\mu_n}(k^\perp)u(k_N)\!&=&\!
X^{(n)}_{\mu_1\ldots\mu_n}(k^\perp)u(k_N)\,.
\nn
N^-_{\mu_1\ldots\mu_{n}}(k^\perp)u(k_N)\!&=&\!
i\gamma_5 \gamma_\nu
X^{(n+1)}_{\nu\mu_1\ldots\mu_{n}}(k^\perp)u(k_N) \,.
\ee
Here, $u(k_N)$ is the bispinor of the final--state nucleon.

In the c.m.s. of the reaction  this amplitude can be rewritten as
\be
&&A_{\pi N}=\omega^*\left [G(s,t)+H(s,t)i(\vec \sigma \vec n)
\right ]\omega' \;,
\nonumber \\
&&G(s,t)=\sum\limits_L \big [(L\!+\!1)F_L^+(s)- L F_L^-(s)\big ]
P_L(z) \;, \nonumber \\
&&H(s,t)=\sum\limits_L \big
[F_L^+(s)+ F_L^-(s)\big ] P'_L(z) \;,
\label{piN_others}
\ee
where $\omega$ and $\omega'$
are nonrelativistic spinors and $\vec n$ is a unit vector
normal to the decay plane. The $F$-functions are defined as follows:
\be
F^+_L&=&(|\vec k||\vec q|)^L \chi_i\chi_f\;\frac{\alpha_L}{2L\!+\!1}
BW_L^+(s) \;,
\nonumber \\
F^-_L&=&(|\vec k||\vec q|)^L \chi_i\chi_f\;\frac{\alpha_L}{L}
BW_L^-(s)\,, \nn \chi_i&=&\sqrt{\frac{m_N+k_{N0}}{2m_N}}\;, \qquad
\chi_f=\sqrt{\frac{m_N+q_{N0}}{2m_N}}\;,
\ee
where $L=n$ stands for '+' states and $L=n+1$ for '-' states.


\section{The electro production amplitudes}

\subsection{The '+' states}

For the states with $n\ge 1$, three vertices can be constructed of the
spin and orbital momentum operators. For '+' states the vertices are:
\be
&&V^{(1+)\mu}_{\alpha_1\ldots\alpha_n}(k^\perp)=
\gamma^\perp_\mu i\gamma_5
X^{(n)}_{\alpha_1\ldots\alpha_n}(k^\perp) \;,\nonumber\\
&&V^{(2+)\mu}_{\alpha_1\ldots\alpha_n}(k^\perp)=
\gamma_\nu i \gamma_5
X^{(n+2)}_{\mu\nu\alpha_1\ldots\alpha_n}(k^\perp) \nonumber \;,\nn
&&V^{(3+)\mu}_{\alpha_1\ldots\alpha_n}(k^\perp)=
\gamma_\nu i \gamma_5
X^{(n)}_{\nu\alpha_1\ldots\alpha_{n-1}}(k^\perp)
g^\perp_{\mu\alpha_n} \;.
\label{vf_plus}
\ee
The first vertex is constructed using the spin $1/2$ operator and
$L=n$ orbital momentum operator, the second one has $S=3/2$, $L=n+2$
and the third one $S=3/2$ and $L=n$. In case of photoproduction, the
second vertex is reduced to the third one and only two amplitudes
(one for $J=1/2$) are independent.

\subsection{ The '-' states}

For the decay of a '-' state with total spin $J$ into $\gamma N$,
the vertex functions have the form:
\be
&&V^{(1-)\mu}_{\alpha_1\ldots\alpha_{n}}(k^\perp)=
\gamma_\xi\gamma^\perp_\mu
X^{(n+1)}_{\xi\alpha_1\ldots\alpha_{n}}(k^\perp) \;,
\nn
&&V^{(2-)\mu}_{\alpha_1\ldots\alpha_{n}}(k^\perp)=
X^{(n+1)}_{\mu\alpha_1\ldots\alpha_{n}}(k^\perp) \;,
\nonumber \\
&&V^{(3-)\mu}_{\alpha_1\ldots\alpha_{n}}(k^\perp)=
X^{(n-1)}_{\alpha_2\ldots\alpha_{n}}(k^\perp)
g^\perp_{\alpha_1\mu} \;.
\label{vf_minus}
\ee
These vertices are constructed of the spin and orbital momentum
operators with ($S=1/2$, $L=n+1$), ($S=3/2$, $L=n+1$) and
($S=3/2$ and $L=n-1$). As in case of "+" states, the second vertex
provides us the same angular distribution as the third vertex. For the
first and third vertices, the width factors $W^-_{i,j}$ are equal to

\newpage
\subsection{Single meson electro-production}

General structure of the single--meson electro-production amplitude
in c.m.s. of the reaction is given by
\be
J_\mu\!=\!
i {\mathcal F_1}
 \tilde\sigma_\mu\! +&&\!\!{\mathcal F_2} (\vec \sigma \vec q)
\frac{\varepsilon_{\mu i j} \sigma_i k_j}{|\vec k| |\vec q|} \! +\!i
{\mathcal F_3} \frac{(\vec \sigma \vec k)}{|\vec k| |\vec q|} \tilde
q_\mu \!+\!i {\mathcal F_4} \frac{(\vec \sigma \vec q)}{\vec q^2}
\tilde q_\mu\,\nn+&&\!i {\mathcal F_5} \frac{(\vec \sigma \vec
k)}{|\vec k|^2} k_\mu \!+\!i {\mathcal F_6} \frac{(\vec \sigma \vec
q)}{|\vec q||\vec k|} k_\mu\,,
\label{mult_1}
\ee
where $\vec q$ is the momentum of the nucleon in the
$\pi N$ channel and $\vec k$ the momentum of the nucleon in the
$\gamma N$ channel calculated in  the c.m.s. of the reaction. The
$\sigma_i$ are Pauli matrices.
\be
\tilde \sigma_\mu&=&\sigma_\mu-\frac{\vec \sigma \vec k}{|\vec
k|^2}k_\mu \qquad \mu=1,2,3\nn \tilde q_\mu&=&q_\mu-\frac{\vec q
\vec k}{|\vec k|^2}k_\mu=q_\mu-z\,k_\mu\frac{|\vec q|}{|\vec k|}
\ee

The functions ${\mathcal F_i}$ have the following angular dependence:
\be
{\mathcal F_1}(z) &= \sum\limits^{\infty}_{L=0}& [LM^+_L\!+\!E^+_L]
P^{\prime}_{L+1}(z)\! +\! \nn &&[(L\!+\!1)M^-_L\! +\! E^-_L]
P^{\prime}_{L-1}(z), \nn {\mathcal F_2}(z) &=
\sum\limits^{\infty}_{L=1}& [(L+1)M^+_L+LM^-_L] P^{\prime}_{L}(z)
\;, \nn {\mathcal F_3}(z) &= \sum\limits^{\infty}_{L=1}&
[E^+_L-M^+_L] P^{\prime\prime}_{L+1}(z) + [E^-_L + M^-_L]
P^{\prime\prime}_{L-1}(z)\;, \nn {\mathcal F_4} (z) &=
\sum\limits^{\infty}_{L=2}& [M^+_L - E^+_L - M^-_L -E^-_L]
P^{\prime\prime}_{L}(z)\;, \nn {\mathcal F_5} (z) &=
\sum\limits^{\infty}_{L=0}& [(L+1)\mathcal L^+_L\,P^{\prime}_{L+1}(z) -
L\mathcal L^-_L P^{\prime}_{L-1}(z)]\;,\nn {\mathcal F_6} (z) &=
\sum\limits^{\infty}_{L=1}& [L\mathcal L^-_L - (L+1)\mathcal L^+_L] P^{\prime}_{L}(z)
\label{mult_2}
\ee
Here $L$ corresponds to the orbital angular momentum in the
$\pi N$ system,
$P_L(z)$, $P'_L(z)$, $P''_L(z)$ are Legendre polynomials and thier
derivatives,
$z=(\vec k\vec q)/(|\vec  k||\vec q|)$, and $E^\pm_L$ and $M^\pm_L$ are
electric and magnetic multipoles describing transitions to
states with $J=L\pm 1/2$.

The single-meson production amplitude via
the intermediate resonance with $J\!=\!n\!+\!1/2$
(we take pion photoproduction as an example)
has the general form:
\be
A^{i\pm}&&=g_{\pi N}(s)\bar u(q_N)
\tilde N^\pm_{\alpha_1\ldots\alpha_n}(q^\perp)\times
\nn
&&\frac{F^{\alpha_1\ldots\alpha_n}_{\beta_1\ldots\beta_n}}
{M^2-s-iM\Gamma_{tot}}
V^{(i\pm)\mu}_{\beta_1\ldots\beta_n}(k^\perp)
u(k_N) g_i(s) \varepsilon_\mu \;.~~
\label{smp_amp}
\ee
Here, $q_N$ and $k_N$ are the momenta of the nucleon in the $\pi N$
and $\gamma N$ channel and $q^\perp$ and $k^\perp$ are
the components of relative momenta which are orthogonal to
the total momentum of the resonance. The index $i$ lists the
$\gamma N$ vertices given in (\ref{vf_plus}), (\ref{vf_minus}).

\subsection{Positive sector}

For the positive amplitudes $L=n$. The spin $\frac 12$ amplitude
has the structure:
\be
A^{1+}_\mu&=&\bar u(q_N)
 X^{(n)}_{\alpha_1\ldots\alpha_n}(q^\perp)\nn
&& F^{\alpha_1\ldots\alpha_n}_{\beta_1\ldots\beta_n}
\gamma_\mu i\gamma_5 X^{(n)}_{\beta_1\ldots\beta_n}(k^\perp)
u(k_N)
\label{a1_pos}
\ee

\be
{\mathcal F^{1+}_1}&=&\lambda_n\,P'_{n+1} \nn
{\mathcal F^{1+}_2}&=&\lambda_n\,P'_{n}\nn
{\mathcal F^{1+}_3}&=&0 \nn
{\mathcal F^{1+}_4}&=&0 \nn
{\mathcal F^{1+}_5}&=&+\lambda_n\,P'_{n+1}\nn
{\mathcal F^{1+}_6}&=&-\lambda_n\,P'_{n}
\ee
where
\be
\lambda_n=\frac{\alpha^{(n)}}{2n+1}(|\vec k||\vec q|)^n\chi_i\chi_f
\ee
\be
\alpha^{(n)}=\prod\limits_{j=1}^{n}\frac{2j-1}{j}\qquad
\alpha^{(0)}=1
\label{alpha}
\ee

Therefore
\be
E_n^{1+}\!=M_n^{1+}\!=\mathcal L_n^{1+}\!=\frac{\lambda_n}{n\!+\!1}\,
\ee

The second ($S=\frac 32$) amplitude has the structure:
\be
A^{3+}_\mu&=&\bar u(q_N)
 X^{(n)}_{\alpha_1\ldots\alpha_n}(q^\perp)\nn
&& F^{\alpha_1\ldots\alpha_n}_{\mu\beta_2\ldots\beta_n}
\gamma_\chi i\gamma_5 X^{(n)}_{\chi\beta_2\ldots\beta_n}(k^\perp)
u(k_N)
\label{a2_pos}
\ee

\be
{\mathcal F^{3+}_1}&=&0 \nn {\mathcal
F^{3+}_2}&=-&\frac{\lambda_n}{n}\,P'_{n}\nn {\mathcal
F^{3+}_3}&=&\frac{\lambda_n}{n}\,P''_{n+1}\nn {\mathcal
F^{3+}_4}&=-&\frac{\lambda_n}{n}\,P''_{n} \nn {\mathcal
F^{3+}_5}&=&+\lambda_n\,P'_{n+1}\nn {\mathcal
F^{3+}_6}&=&-\lambda_n\,P'_{n}
\ee

Therefore
\be
E_n^{3+}\!=\mathcal L_n^{3+}\!=-nM_n^{3+}=\frac{\lambda_n}{n\!+\!1}\,
\ee

The third amplitude has the structure:
\be
A^{2+}_\mu&=&\bar u(q_N)
X^{(n)}_{\alpha_1\ldots\alpha_n}(q^\perp)\nn
&&F^{\alpha_1\ldots\alpha_n}_{\beta_1\ldots\beta_n} \gamma_\chi
i\gamma_5 X^{(n+2)}_{\mu\chi\beta_1\ldots\beta_n}(k^\perp) u(k_N)
\label{a3_pos}
\ee

\be
{\mathcal F^{2+}_1}&=&\xi_n\,P'_{n+1} \nn {\mathcal F^{2+}_2}&=&0\nn
{\mathcal F^{2+}_3}&=&\xi_n\,P''_{n+1}\nn {\mathcal
F^{2+}_4}&=-&\xi_n\,P''_{n} \nn {\mathcal
F^{2+}_5}&=-&\xi_n(n\!+\!2)\,P'_{n+1}\nn {\mathcal
F^{2+}_6}&=&\xi_n(n\!+\!2)\,P'_{n}
\ee
where
\be
\xi_n=\frac{|\vec k|^2(2n\!+\!1)}{(n\!+\!2)(n\!+\!1)}\lambda_n=
\frac{\alpha^{(n)}|\vec k|^{n+2}|\vec
q|^n}{(n\!+\!2)(n\!+\!1)}\chi_i\chi_f
\ee
Therefore
\be
E_n^{2+}\!=\xi^n\qquad M_n^{2+}\!=0\qquad
\mathcal L_n^{2+}\!=-\xi_n\frac{n\!+\!2}{n\!+\!1},
\ee

\subsection{Negative sector}

For the negative amplitudes $L=n+1$. The spin $\frac 12$ amplitude
has the structure:
\be
A^{1-}_\mu&=&\bar u(q_N)
 X^{(n+1)}_{\alpha_1\ldots\alpha_n\nu}(q^\perp)\gamma_\nu i\gamma_5 \nn
&& F^{\alpha_1\ldots\alpha_n}_{\beta_1\ldots\beta_n}
\gamma_\xi\gamma_\mu X^{(n+1)}_{\xi\beta_1\ldots\beta_n}(k^\perp)
u(k_N)
\label{a1_neg}
\ee

\be
{\mathcal F^{1-}_1}&=&-\zeta_{n+1}\,P'_{n} \nn
{\mathcal F^{1-}_2}&=&-\zeta_{n+1}\,P'_{n+1}\nn
{\mathcal F^{1-}_3}&=&0 \nn
{\mathcal F^{1-}_4}&=&0 \nn
{\mathcal F^{1-}_5}&=&+\zeta_{n+1}\,P'_{n}\nn
{\mathcal F^{1-}_6}&=&-\zeta_{n+1}\,P'_{n+1}
\ee
where
\be
\zeta_{n+1}=\frac{\alpha^{(n+1)}}{n\!+\!1}(|\vec k||\vec
q|)^{n+1}\chi_i\chi_f
\ee

Therefore
\be
-M_{n+1}^{1-}\!=E_{n+1}^{1-}\!=\mathcal L_{n+1}^{1-}\!=\frac{\zeta_{n+1}}{n\!+\!1}\,
\ee

For the third negative amplitude (spin $\frac 32$):
\be
A^{3-}_\mu&=&\bar u(q_N)
 X^{(n+1)}_{\alpha_1\ldots\alpha_n\nu}(q^\perp)\gamma_\nu i\gamma_5 \nn
&& F^{\alpha_1\ldots\alpha_n}_{\mu\beta_2\ldots\beta_n}
X^{(n-1)}_{\beta_2\ldots\beta_n}(k^\perp)
u(k_N)
\label{a3_neg}
\ee

\be
{\mathcal F^{3-}_1}&=&\varrho_{n-1}\,P'_{n} \nn
{\mathcal F^{3-}_2}&=&0\nn
{\mathcal F^{3-}_3}&=&\varrho_{n-1}\,P''_n \nn
{\mathcal F^{3-}_4}&=-&\varrho_{n-1}\,P''_{n+1} \nn
{\mathcal F^{3-}_5}&=&n\varrho_{n-1}\,P'_{n}\nn
{\mathcal F^{3-}_6}&=&n\varrho_{n-1}\,P'_{n+1}
\ee
where
\be
\varrho_{n-1}=\frac{\alpha^{(n-1)}}{n(n\!+\!1)}|\vec k|^{n-1}|\vec
q|^{n+1}\chi_i\chi_f
\ee

Therefore
\be
M_{n+1}^{3-}\!=0\qquad E_{n+1}^{3-}\!=\varrho_{n-1}
\qquad \mathcal L_{n+1}^{3-}\!=\varrho_{n-1}\frac{n}{n\!+\!1}\,
\ee

For the second amplitude from negative sector:
\be
A^{2-}_\mu&=&\bar u(q_N)
 X^{(n+1)}_{\alpha_1\ldots\alpha_n\nu}(q^\perp)\gamma_\nu i\gamma_5 \nn
&& F^{\alpha_1\ldots\alpha_n}_{\beta_1\ldots\beta_n}
X^{(n+1)}_{\mu\beta_1\ldots\beta_n}(k^\perp)
u(k_N)
\label{a2_neg}
\ee

\be
{\mathcal F^{2-}_1}&=&\Delta_{n}\,P'_{n} \nn
{\mathcal F^{2-}_2}&=&0\nn
{\mathcal F^{2-}_3}&=&\Delta_{n}\,P''_n \nn
{\mathcal F^{2-}_4}&=-&\Delta_{n}\,P''_{n+1} \nn
{\mathcal F^{2-}_5}&=-&(n+1)\Delta_{n}\,P'_{n}\nn
{\mathcal F^{2-}_6}&=&(n+1)\Delta_{n}\,P'_{n+1}
\ee
where
\be
\Delta_{n}=\frac{\alpha^{(n)}}{(n\!+\!1)^2} (|\vec k||\vec
q|)^{n+1}\chi_i\chi_f
\ee

Therefore
\be
M_{n+1}^{2-}\!=0\qquad E_{n+1}^{2-}\!=-\Delta_n
\qquad \mathcal L_{n+1}^{2-}\!=-\Delta_n\,
\ee

Remember that $\chi_i\!=\!m_N\!+\!q_{N0}$ and $\chi_f\!=\!m_N\!+\!k_{N0}$.
For the '-' states, where $L\!=\!n\!+\!1$, the corresponding
equations are
\be
E_L^{-(\frac12)}&=&- \sqrt{\chi_i\chi_f}\; \frac{\alpha^{(L)}}{L^2}
\frac{g_{\pi N}(|\vec k||\vec q|)^L g_1(s)}
{M^2-s-iM\Gamma_{tot}}\;, \nn
M_L^{-(\frac12)}&=&-E_L^{-(\frac12)}\;, \nn
E_L^{-(\frac32)}&=&-\frac{\alpha^{(L-2)}}{(L\!-\!1)L}
\sqrt{\chi_i\chi_f}\; \frac{g_{\pi N}|\vec k|^{L-2}|\vec q|^L
g_3(s)} {M^2-s-iM\Gamma_{tot}}\;, \nn M_L^{-(\frac32)}&=&0  \;.
\label{mpd_n2}
\ee
These formulae are different from the correspondent expressions
given in \cite{Anisovich:2007zz} by the factor $(-1)^n$ which enters
now in the resonance propagator. All other formulae given in
\cite{Anisovich:2007zz} for the single meson photoproduction are not
changed due to this redefinition.

The second ($S=\frac 32$) amplitude has the structure:
\be
A^{2+}_\mu=\bar u(q_N)
 X^{(n)}_{\alpha_1\ldots\alpha_n}(q^\perp)F^{\alpha_1\ldots\alpha_n}_{\mu\beta_2\ldots\beta_n}
\gamma_\chi i\gamma_5 X^{(n)}_{\chi\beta_2\ldots\beta_n}(k^\perp)
u(k_N)\nn
\label{a2p_pos}
\ee

\section{The gauge invariant vertices}

\subsection{The '+' states}

Here we have three vertices.
\be
&&V^{(1+)\mu}_{\alpha_1\ldots\alpha_n}(k^\perp)=
\gamma^\perp_\mu i\gamma_5
X^{(n)}_{\alpha_1\ldots\alpha_n}(k^\perp) \;,\nonumber\\
&&V^{(2+)\mu}_{\alpha_1\ldots\alpha_n}(k^\perp)=
\gamma_\nu i \gamma_5
X^{(n+2)}_{\mu\nu\alpha_1\ldots\alpha_n}(k^\perp) \nonumber \;,\nn
&&V^{(3+)\mu}_{\alpha_1\ldots\alpha_n}(k^\perp)=
\gamma_\nu i \gamma_5
X^{(n)}_{\nu\alpha_1\ldots\alpha_{n-1}}(k^\perp)
g^\perp_{\mu\alpha_n} \;.
\label{vf_plus1}
\ee
The vertices (1)and (3) are used to fit
the photo-production reactions. Let us consider the vertex 2 with a
propagator of the baryon state:
\be
&&F^{\alpha_1\ldots\alpha_n}_{\beta_1\ldots
\beta_n}V^{(2+)\mu}_{\alpha_1\ldots\alpha_n}(k^\perp)=
F^{\alpha_1\ldots\alpha_n}_{\beta_1\ldots
\beta_n}\gamma_\nu i
\gamma_5 X^{(n+2)}_{\mu\nu\alpha_1\ldots\alpha_n}(k^\perp)=\nn
&&F^{\alpha_1\ldots\alpha_n}_{\beta_1\ldots
\beta_n}\gamma_\nu i\gamma_5 \alpha^{(n+2)}
\Big (k^\perp_\mu k^\perp_\nu k^\perp_{\alpha_1}\ldots
k^\perp_{\alpha_n}-\frac{k_\perp^2}{2n+3}\times
\nn&&\big(g^\perp_{\mu\nu}k^\perp_{\alpha_1}\ldots
k^\perp_{\alpha_n} +g^\perp_{\mu\alpha_1}k^\perp_\nu
k^\perp_{\alpha_2}\ldots k^\perp_{\alpha_n}
+g^\perp_{\nu\alpha_1}k^\perp_\mu k^\perp_{\alpha_2}\ldots
k^\perp_{\alpha_n}\nn &&+g^\perp_{\alpha_1\alpha_2}k^\perp_\mu
k^\perp_\nu k^\perp_{\alpha_3}\ldots k^\perp_{\alpha_n}+\ldots\big
)+ \frac{k_\perp^4}{(2n+3)(2n+1)}\times \nn&&\big
(g^\perp_{\mu\nu}g^\perp_{\alpha_1\alpha_2} k^\perp_{\alpha_3}\ldots
k^\perp_{\alpha_n} g^\perp_{\mu\alpha_1}g^\perp_{\alpha_2\alpha_3}
k^\perp_{\nu}k^\perp_{\alpha_4}\ldots k^\perp_{\alpha_n}+\nn
&&g^\perp_{\mu\alpha_1}g^\perp_{\nu\alpha_2}
k^\perp_{\alpha_3}\ldots k^\perp_{\alpha_n}+\ldots\big
)\ldots\Big)~~~~~~~~~~~
\label{v2_plus}
\ee
Taking into account that
\be
F^{\alpha_1\ldots\alpha_n}_{\beta_1\ldots\beta_n}g_{\alpha_i\alpha_j}=0\qquad
F^{\alpha_1\ldots\alpha_n}_{\beta_1\ldots\beta_n}\gamma_\nu g_{\nu\alpha_j}=0
\ee
we obtain that this vertex can be written as:
\be
&&F^{\alpha_1\ldots\alpha_n}_{\beta_1\ldots
\beta_n}V^{(2+)\mu}_{\alpha_1\ldots\alpha_n}(k^\perp)=\frac{\alpha^{(n+2)}}{\alpha^{(n)}}
F^{\alpha_1\ldots\alpha_n}_{\beta_1\ldots
\beta_n}\times\nn&&\big (\hat k^\perp i \gamma_5 k^\perp_\mu X^{(n)}_{\alpha_1\ldots\alpha_n}(k^\perp)
\nn&& -\frac{k_\perp^2}{2n+3}\big
(V^{(1+)\mu}_{\alpha_1\ldots\alpha_n}(k^\perp)+
n\;V^{(3+)\mu}_{\alpha_1\ldots\alpha_n}(k^\perp)\big)
\label{tilde_v2}
\ee
It means that instead of
$V^{(2+)\mu}_{\alpha_1\ldots\alpha_n}(k^\perp)$ one can use the
vertex:
\be
\tilde V^{(2+)\mu}_{\alpha_1\ldots\alpha_n}(k^\perp)=
\hat k^\perp i \gamma_5 k^\perp_\mu X^{(n)}_{\alpha_1\ldots\alpha_n}(k^\perp)
\ee

Let us calculate the convolution of the vertices with photon momentum. Remember:
\be
k^\perp_\mu=\frac 12 (k^N_\nu-k^\gamma_\nu)\;g^\perp_{\mu\nu}
\ee
Thus:
\be
k^\gamma_\nu g^\perp_{\mu\nu}=-k^\perp_\mu\quad k^\gamma_\nu k^\perp_\nu=-k_\perp^2
\ee
Therefore:
\be
V^{(1+)\mu}_{\alpha_1\ldots\alpha_n}(k^\perp)k^\gamma_\mu&=&-\hat k^\perp i \gamma_5
X^{(n)}_{\alpha_1\ldots\alpha_n}(k^\perp)\nonumber \\
\tilde V^{(2+)\mu}_{\alpha_1\ldots\alpha_n}(k^\perp)k^\gamma_\mu&=&-\hat k^\perp i \gamma_5
 X^{(n)}_{\alpha_1\ldots\alpha_n}(k^\perp) k^2_\perp
\ee

It means that the gauge invariant operator can be made as:
\be
V^{G(1+)\mu}_{\alpha_1\ldots\alpha_n}(k^\perp)=V^{(1+)\mu}_{\alpha_1\ldots\alpha_n}(k^\perp)-
\frac{1}{k_\perp^2} \tilde V^{(2+)\mu}_{\alpha_1\ldots\alpha_n}(k^\perp)
\ee
Which leads to a very simple expression:
\be
V^{G(1+)\mu}_{\alpha_1\ldots\alpha_n}(k^\perp)=\gamma^{\perp\perp}_\mu i\gamma_5
X^{(n)}_{\alpha_1\ldots\alpha_n}(k^\perp)
\ee
The third vertex should be considered with propagator of the
resonance. Thus the convolution with photon momentum:
\be
F^{\alpha_1\ldots\alpha_n}_{\beta_1\ldots\beta_n}V^{(3+)\mu}_{\alpha_1\ldots\alpha_n}(k^\perp)k^\gamma_\mu=
-F^{\alpha_1\ldots\alpha_n}_{\beta_1\ldots\beta_n}\gamma_\nu i
\gamma_5 X^{(n)}_{\nu\alpha_2\ldots\alpha_n}k^\perp_{\alpha_1}
\nonumber \\=-F^{\alpha_1\ldots\alpha_n}_{\beta_1\ldots\beta_n}
\hat k^\perp
i \gamma_5 X^{(n)}_{\alpha_1\ldots\alpha_n}~~~~~~~~~~~
\ee
Which coincides with vertex (1). It means that gauge invariant combination could be
\be
V^{(3+)\mu}_{\alpha_1\ldots\alpha_n}(k^\perp)-V^{(1+)\mu}_{\alpha_1\ldots\alpha_n}(k^\perp)
\ee
which is used in some of articles.
However for us it is easier to use another combination:
\be
V^{G(3+)\mu}_{\alpha_1\ldots\alpha_n}(k^\perp)=
V^{(3+)\mu}_{\alpha_1\ldots\alpha_n}(k^\perp)-
\frac{1}{k_\perp^2} \tilde V^{(2+)\mu}_{\alpha_1\ldots\alpha_n}(k^\perp)
\ee
In the presence of the resonance propagator it can be rewritten as:
\be
V^{G(3+)\mu}_{\alpha_1\ldots\alpha_n}(k^\perp)=\gamma_\nu i \gamma_5
X^{(n)}_{\nu\alpha_1\ldots\alpha_{n-1}}(k^\perp)
g^{\perp\perp}_{\mu\alpha_n}
\ee

The second vertex can be written in the gauge invariant form using the property:
\be
k^\perp_\nu\big (g_{\mu\nu}-\frac{k^\gamma_\nu P_\mu}{(Pk^\gamma)}\big )=
k^\perp_\mu+\frac{k^2_\perp}{(Pk^\gamma)}P_\mu
\ee
Taking into account that:
\be
k^2_\perp=(k^\gamma)^2-\frac{(k^\gamma P)^2}{P^2} \qquad k_\mu^\perp=-k^\gamma_\mu+
\frac{(k^\gamma P)}{P^2}P_\mu
\ee
We obtain:

\be
k^\perp_\nu\big(g_{\mu\nu}-
\frac{k^\gamma_\nu P_\mu}{(Pk^\gamma)}\big )=-k^\gamma_\mu+\frac{(k^\gamma)^2}{(Pk^\gamma)}P_\mu
\ee

This is a gauge invariant vertex. Due to orthogonality of the photon
momentum to its polarization vector one can reduce it to:
\be
V^{G(2+)\mu}_{\alpha_1\ldots\alpha_n}(k^\perp)= \hat k^\perp i
\gamma_5 P_\mu
X^{(n)}_{\alpha_1\ldots\alpha_n}(k^\perp)\frac{(k^\gamma)^2}{(Pk^\gamma)}
\ee

Thus we can use:
\be
V^{G(1+)\mu}_{\alpha_1\ldots\alpha_n}(k^\perp)&=&\gamma^{\perp\perp}_\mu
i\gamma_5 X^{(n)}_{\alpha_1\ldots\alpha_n}(k^\perp) \nonumber \\
V^{G(2+)\mu}_{\alpha_1\ldots\alpha_n}(k^\perp)&=& \hat k^\perp i
\gamma_5
X^{(n)}_{\alpha_1\ldots\alpha_n}(k^\perp)\left (P_\mu\frac{(k^\gamma)^2}{(Pk^\gamma)}-
k^\gamma_\mu\right)
\nonumber \\
V^{G(3+)\mu}_{\alpha_1\ldots\alpha_n}(k^\perp)&=&\gamma_\nu i \gamma_5
X^{(n)}_{\nu\alpha_1\ldots\alpha_{n-1}}(k^\perp)
g^{\perp\perp}_{\mu\alpha_n}
\ee

Let us calculate the structure for $\tilde V^{(2+)}$:
\be
\tilde A^{2+}_\mu&=&\bar u(q_N)
X^{(n)}_{\alpha_1\ldots\alpha_n}(q^\perp)\nn
&&F^{\alpha_1\ldots\alpha_n}_{\beta_1\ldots\beta_n} \hat k^\perp i
\gamma_5 k^\perp_\mu X^{(n)}_{\alpha_1\ldots\alpha_n}(k^\perp)
\label{a2_tilde}
\ee
Using eq.(~\ref{tilde_v2}) we obtain:
\be
\tilde F^{2+}_i=F^{2+}_i \frac{\alpha^{(n)}}{\alpha^{(n+2)}}+
\frac{k_\perp^2}{2n+3}\big (F^{1+}_i +n\;F^{3+}_i\big )
\ee
Taking into account:
\be
\xi_n\;\frac{\alpha^{(n)}}{\alpha^{(n+2)}}=\frac{-k_\perp^2}{2n+3}
\ee

\be
&{\tilde\mathcal F^{2+}_1}=0 \qquad &{\tilde \mathcal F^{2+}_2}=0\nn
&{\tilde\mathcal F^{2+}_3}=0\qquad &{\tilde\mathcal F^{2+}_4}=0 \nn
&{\tilde\mathcal F^{2+}_5}=k^2_\perp\lambda_n\,P'_{n+1}\qquad
&{\tilde\mathcal F^{2+}_6}=-k^2_\perp\lambda_n\,P'_{n}
\ee
we obtain:
\be
E_n^{2+}\!=M_n^{2+}\!=0\qquad \mathcal L_n^{1+}=\frac{k^2_\perp}{n+1}\lambda_n\!
\ee
Then the first vertex will be:
\be
A^{G(1+)}_\mu&=&\bar u(q_N)
 X^{(n)}_{\alpha_1\ldots\alpha_n}(q^\perp)
 F^{\alpha_1\ldots\alpha_n}_{\beta_1\ldots\beta_n}
V^{G(1+)}_{\beta_1\ldots\beta_n}(k^\perp) u(k_N) \nn
\label{ta1_pos}
\ee
Then we obtain:
\be
{\mathcal F^{G(1+)}_1}&=&\lambda_n\,P'_{n+1} \nn {\mathcal
F^{G(1+)}_2}&=&\lambda_n\,P'_{n}\nn {\mathcal
F^{G(1+)}_3}&=&{\mathcal F^{G(1+)}_4}=0 \nn {\mathcal
F^{G(1+)}_5}&=& {\mathcal F^{G(1+)}_6}=0
\ee
Therefore
\be
E_n^{1+}\!=M_n^{1+}\!=\frac{\lambda_n}{n\!+\!1}\,\qquad \mathcal L_n^{1+}=0\!
\ee

The second ($S=\frac 32$) amplitude has the structure:
\be
A^{G(3+)}_\mu=\bar u(q_N)
 X^{(n)}_{\alpha_1\ldots\alpha_n}(q^\perp)
 F^{\alpha_1\ldots\alpha_n}_{\mu\beta_2\ldots\beta_n}
V^{G(3+)}_{\beta_1\ldots\beta_n}(k^\perp) u(k_N)\nn
\label{ta3_pos}
\ee
and
\be
{\mathcal F^{G(3+)}_1}&=&0 \nn {\mathcal
F^{G(3+)}_2}&=-&\frac{\lambda_n}{n}\,P'_{n}\nn {\mathcal
F^{G(3+)}_3}&=&\frac{\lambda_n}{n}\,P''_{n+1}\nn {\mathcal
F^{G(3+)}_4}&=-&\frac{\lambda_n}{n}\,P''_{n} \nn {\mathcal
F^{G(3+)}_5}&=&{\mathcal
F^{G(3+)}_6}=0
\ee

Therefore
\be
E_n^{3+}\!=-nM_n^{3+}=\frac{\lambda_n}{n\!+\!1}\,\qquad
\mathcal L_n^{3+}=0\!
\ee

Thus we obtain a behavior of the CGLN functions without
any kinematical problems.

\subsection{ The '-' states}

For the decay of a '-' state with total spin $J$ into $\gamma N$,
the vertex functions have the form:
\be
&&V^{(1-)\mu}_{\alpha_1\ldots\alpha_{n}}(k^\perp)=
\gamma_\xi\gamma^\perp_\mu
X^{(n+1)}_{\xi\alpha_1\ldots\alpha_{n}}(k^\perp) \;,
\nn
&&V^{(2-)\mu}_{\alpha_1\ldots\alpha_{n}}(k^\perp)=
X^{(n+1)}_{\mu\alpha_1\ldots\alpha_{n}}(k^\perp) \;,
\nonumber \\
&&V^{(3-)\mu}_{\alpha_1\ldots\alpha_{n}}(k^\perp)=
X^{(n-1)}_{\alpha_2\ldots\alpha_{n}}(k^\perp)
g^\perp_{\alpha_1\mu} \;.
\label{vf1_minus}
\ee

If one follows this idea we obtain the following vertices:
\be
&&V^{G(1-)\mu}_{\alpha_1\ldots\alpha_{n}}(k^\perp)=
\gamma_\xi\gamma^{\perp\perp}_\mu
X^{(n+1)}_{\xi\alpha_1\ldots\alpha_{n}}(k^\perp) \;,
\nn
&&V^{G(2-)\mu}_{\alpha_1\ldots\alpha_{n}}(k^\perp)=
X^{(n)}_{\alpha_1\ldots\alpha_{n}}(k^\perp)\left (P_\mu\frac{(k^\gamma)^2}{(Pk^\gamma)}-
k^\gamma_\mu\right),
\nn
&&V^{G(3-)\mu}_{\alpha_1\ldots\alpha_{n}}(k^\perp)=
X^{(n-1)}_{\alpha_2\ldots\alpha_{n}}(k^\perp)
g^{\perp\perp}_{\alpha_1\mu} \;.
\label{vf2_minus}
\ee

If one follows this idea we obtain the following expressions. For
the first vertex:
\be
{\mathcal F^{G(1-)}_1}&=&-\zeta_{n+1}\,P'_{n} \nn {\mathcal
F^{G(1-)}_2}&=&-\zeta_{n+1}\,P'_{n+1}\nn {\mathcal F^{G(1-)}_3}&=&0 \nn
{\mathcal F^{G(1-)}_4}&=&0 \nn {\mathcal F^{G(1-)}_5}&=&0 \nn {\mathcal
F^{G(1-)}_6}&=&0
\ee
where
\be
\zeta_{n+1}=\frac{\alpha^{(n+1)}}{n\!+\!1}(|\vec k||\vec
q|)^{n+1}\chi_i\chi_f
\ee
And therefore for this vertex: Therefore
\be
-M_{n+1}^{1-}\!=E_{n+1}^{1-}\!=\frac{\zeta_{n+1}}{n\!+\!1}\,\qquad
\mathcal L_{n+1}^{1-}\!=0
\ee

The third vertex:
\be
{\mathcal F^{G(3-)}_1}&=&\varrho_{n-1}\,P'_{n} \nn {\mathcal
F^{G(3-)}_2}&=&0\nn {\mathcal F^{G(3-)}_3}&=&\varrho_{n-1}\,P''_n \nn
{\mathcal F^{G(3-)}_4}&=&-\varrho_{n-1}\,P''_{n+1} \nn {\mathcal
F^{G(3-)}_5}&=&0\nn {\mathcal F^{G(3-)}_6}&=&0
\ee
where
\be
\varrho_{n-1}=\frac{\alpha^{(n-1)}}{n(n\!+\!1)}|\vec k|^{n-1}|\vec
q|^{n+1}\chi_i\chi_f
\ee
\be
M_{n+1}^{3-}\!=0\qquad E_{n+1}^{3-}\!=\varrho_{n-1} \qquad \mathcal
L_{n+1}^{3-}\!=0\,
\ee
For the second vertex we obtain:
\be
{\mathcal F^{G(2-)}_1}&=&0\nn
{\mathcal F^{G(2-)}_2}&=&0\nn
{\mathcal F^{G(2-)}_3}&=&0\nn
{\mathcal F^{G(2-)}_4}&=&0\nn
{\mathcal F^{G(2-)}_5}&=&-(n+1)\Delta_{n}\,P'_{n}\nn
{\mathcal F^{G(2-)}_6}&=&(n+1)\Delta_{n}\,P'_{n+1}
\ee
where
\be
\Delta_{n}=\frac{\alpha^{(n)}}{(n\!+\!1)^2} (|\vec k||\vec
q|)^{n+1}\chi_i\chi_f
\ee

Therefore
\be
M_{n+1}^{2-}\!=E_{n+1}^{2-}\!=0
\qquad \mathcal L_{n+1}^{2-}\!=\Delta_n\,
\ee

\section{The connection with other vertex definitions}

In the article \cite{Krivoruchenko:2001jk} the electro-production amplitudes were introduced as
\be
A_{K\mu}^{(\pm)i}=\bar
u_{\beta_1\ldots\beta_n}(P)\Gamma^{(\pm)i}_{\beta_1\ldots\beta_n\mu}u(k_1)
\label{a_kf}
\ee
where for the '+' sector:
\be
\Gamma^{(+)1}_{\beta_1\ldots\beta_n\mu}&=&
\sqrt{s}\big(q_{\beta_1}\gamma_\mu-\hat q
g_{\beta_1\mu}\big)q_{\beta_2}\ldots q_{\beta_n}i\gamma_5 \nn
\Gamma^{(+)2}_{\beta_1\ldots\beta_n\mu}&=& \big(q_{\beta_1}\tilde
P_\mu-(q\tilde P) g_{\beta_1\mu}\big )q_{\beta_2}\ldots
q_{\beta_n}i\gamma_5 \nn \Gamma^{(+)3}_{\beta_1\ldots\beta_n\mu}&=&
\big(q_{\beta_1}q_\mu-q^2 g_{\beta_1\mu}\big )q_{\beta_2}\ldots
q_{\beta_n}i\gamma_5
\ee
Here $q$ was used in \cite{Krivoruchenko:2001jk} to define the
vector of the virtual photon and therefore $q\equiv k^\gamma$. The
vector $\tilde P_\mu=\frac 12 (P+k_1)_\mu=P_\mu-q_\mu/2$. We also
use our definition of $\gamma_5$ matrix
$\gamma_5=\gamma_0\gamma_1\gamma_2\gamma_3$ which is differ from
\cite{Krivoruchenko:2001jk} by the factor $i$. Another difference is
that in the article \cite{Krivoruchenko:2001jk} the normalization of
the resonance polarization vectors is defined as:
\be
&&\bar u(k_1)u(k_1)=2m\nn&& \bar
u_{\beta_1\ldots\beta_n}(P)u_{\beta_1\ldots\beta_n}(P)=(-1)^n2\sqrt
s
\ee
while our definition is:
\be
&&\bar u(k_1)u(k_1)=1\nn&& \bar
u_{\beta_1\ldots\beta_n}(P)u_{\beta_1\ldots\beta_n}(P)=(-1)^n
\ee
It means that our amplitudes are different by the coefficient
\be
N=\frac{1}{2\sqrt{m\sqrt{s}}}
\ee
However we introduced in our propagator an additional factor
$(-1)^n$. The orbital momentum operator in our definition depends on
the relative momentum between nucleon and photon. It can be
rewritten through the relative momentum of the photon and nucleon
as:
\be
X^{(n)}_{\alpha_1\ldots\alpha_n}(k^\perp)=(-1)^n
X^{(n)}_{\alpha_1\ldots\alpha_n}(q^\perp)
\ee
Therefore we have two factors $(-1)^n$ which compensate each
another. If one use definitions of the polarization vectors from
\cite{Krivoruchenko:2001jk} our vertices can be written as:
\be
\tilde
V^{G(1+)\mu}_{\beta_1\ldots\beta_n}(k^\perp)&=&N\gamma^{\perp\perp}_\mu
i\gamma_5 X^{(n)}_{\beta_1\ldots\beta_n}(q^\perp) \nonumber \\
\tilde V^{G(2+)\mu}_{\beta_1\ldots\beta_n}(k^\perp)&=& -N\hat
q^\perp i \gamma_5 X^{(n)}_{\beta_1\ldots\beta_n}(q^\perp)\left
(P_\mu\frac{q^2}{(Pq)}- q_\mu\right)
\nonumber \\
\tilde V^{G(3+)\mu}_{\beta_1\ldots\beta_n}(k^\perp)&=&N\gamma_\nu i
\gamma_5 X^{(n)}_{\nu\beta_2\ldots\beta_{n}}(q^\perp)
g^{\perp\perp}_{\mu\beta_1}
\ee
If one use the definition of momenta and polarization vectors like
in the \cite{Krivoruchenko:2001jk}.

The corresponding amplitudes can be expressed as:
\be
A_\mu^{i\pm}=\bar
u_{\beta_1\ldots\beta_n}(P)V^{G(i\pm)\mu}_{\beta_1\ldots\beta_n}u(k_1)
\ee
where bispinors are taken as defined in \cite{Krivoruchenko:2001jk}.
The relation between these amplitudes (currents) and amplitudes
given in (II.1) of article \cite{Krivoruchenko:2001jk} are:
\be
A_\mu^{1+}&=&N\frac{\alpha^{(n)}}{\sqrt{s}}\Big [
A_{K\mu}^{(+)1}\nn&+&\frac{\chi}{q_\perp^2}\big(A_{K\mu}^{(+)2}(\sqrt
s-m)- \frac 12(\sqrt s+m) A_{K\mu}^{(+)3}\big )\Big ]\nn
A_\mu^{2+}&=&N\frac{\alpha^{(n)}\chi}{(Pq)}\Big [
-q^2A_{K\mu}^{(+)2}+ \frac{s-m^2}{2}A_{K\mu}^{(+)3}\Big ]\\
A_\mu^{3+}&=&N\frac{\alpha^{(n)}\chi}{q_\perp^2}
\Big [
\frac{(Pq)}{s}\big(A_{K\mu}^{(+)2}+\frac 12
A_{K\mu}^{(+)3}\big)-A_{K\mu}^{(+)3}\Big ]
\nonumber
\ee
Here
\be
\chi=m+\sqrt{s}-\frac{(Pq)}{\sqrt{s}}=m+\frac{(Pk_1)}{\sqrt{s}}
\ee
and
\be
q_\perp^2=k_\perp^2=q^2-\frac{(Pq)^2}{s}
\ee

For the '-' sector $J^P=1/2^+,3/2^-,\ldots$ the vertices in
\cite{Krivoruchenko:2001jk} are defined as:
\be
\Gamma^{(-)1}_{\beta_1\ldots\beta_n\mu}&=&
-\sqrt{s}\big(q_{\beta_1}\gamma_\mu-\hat q
g_{\beta_1\mu}\big)q_{\beta_2}\ldots q_{\beta_n} \nn
\Gamma^{(-)2}_{\beta_1\ldots\beta_n\mu}&=&-\big(q_{\beta_1}\tilde
P_\mu-(q\tilde P) g_{\beta_1\mu}\big )q_{\beta_2}\ldots
q_{\beta_n}\nn \Gamma^{(-)3}_{\beta_1\ldots\beta_n\mu}&=&
-\big(q_{\beta_1}q_\mu-q^2 g_{\beta_1\mu}\big )q_{\beta_2}\ldots
q_{\beta_n}
\ee
As before, our amplitudes can be rewritten as:
\be
V^{G(1-)\mu}_{\alpha_1\ldots\alpha_{n}}(k^\perp)&=&
-N\gamma_\xi\gamma^{\perp\perp}_\mu
X^{(n+1)}_{\xi\alpha_1\ldots\alpha_{n}}(q^\perp) \;, \nn
V^{G(2-)\mu}_{\alpha_1\ldots\alpha_{n}}(k^\perp)&=&
NX^{(n)}_{\alpha_1\ldots\alpha_{n}}(q^\perp)\left
(P_\mu\frac{q^2}{(Pq)}- q_\mu\right), \nn
V^{G(3-)\mu}_{\alpha_1\ldots\alpha_{n}}(q^\perp)&=&
-NX^{(n-1)}_{\alpha_2\ldots\alpha_{n}}(k^\perp)
g^{\perp\perp}_{\alpha_1\mu} \;.
\ee
Then we obtain the following relation:
\be
A_\mu^{1-}&=&N\frac{\alpha^{(n\!+\!1)}}{\sqrt{s}}\Big [ \chi
A_{K\mu}^{(-)1}\!-\!(\sqrt{s}\!+\!m)A_{K\mu}^{(-)2}+
\frac{\sqrt{s}\!-\!m}{2} A_{K\mu}^{(-)3} \Big]\nn
A_\mu^{2-}&=&N\frac{\alpha^{(n)}}{(Pq)}\Big [
q^2A_{K\mu}^{(-)2}+ \frac{m^2-s}{2}A_{K\mu}^{(-)3}\Big ]\nn
A_\mu^{3-}&=&N\frac{\alpha^{(n-1)}}{q_\perp^2}\Big [
\frac{(Pq)}{s}\big(A_{K\mu}^{(-)2}+\frac 12
A_{K\mu}^{(-)3}\big)-A_{K\mu}^{(-)3}\Big ]
\ee

For the states with spin 1/2 the situation is more complicated. The
vertices in \cite{Krivoruchenko:2001jk} are defined as:
\be
\Gamma_\mu^{(+)1}&=&\big(q^2\gamma_\mu-\hat q q_\mu\big )i\gamma_5
\nn \Gamma_\mu^{(+)2}&=&\big((\tilde Pq)\gamma_\mu -\hat q \tilde
P_\mu\big )i\gamma_5
\ee
Thus we obtain the following relation:
\be
A_\mu^{1+}&=&\frac{N\chi}{q_\perp^2\sqrt s}\Big [\frac 12
A_{K\mu}^{(+)1}\!-\frac{\sqrt s-m}{\sqrt s+m}A_{K\mu}^{(+)2}\Big]\nn
A_\mu^{2+}&=&\frac{N\chi}{(\sqrt s+m)(Pq)}\Big [q^2A_{K\mu}^{(+)2}+
\frac{m^2-s}{2}A_{K\mu}^{(+)1}\Big ]
\ee
and for the $1/2^+$ state the vertices in
\cite{Krivoruchenko:2001jk} are
\be
\Gamma_\mu^{(-)1}&=&\big(\hat q q_\mu-q^2\gamma_\mu\big ) \nn
\Gamma_\mu^{(-)2}&=&\big(\hat q \tilde P_\mu-(\tilde
Pq)\gamma_\mu\big )
\ee
And we have the following relation:
\be
A_\mu^{1-}&=&\frac{N}{\sqrt s}\Big [-\frac 12
A_{K\mu}^{(-)1}\!+\frac{\sqrt s+m}{\sqrt s-m}A_{K\mu}^{(-)2}\Big]\nn
A_\mu^{2-}&=&\frac{N}{(\sqrt s-m)(Pq)}\Big [q^2A_{K\mu}^{(-)2}+
\frac{m^2-s}{2}A_{K\mu}^{(-)1}\Big ]
\ee

\section{Summary}

We develop the formalism for the analysis of the meson
electro-production reaction. The method is fully based on the
covariant approach used by the Bonn-Gatchina group for the analysis
of the meson photoproduction data and can be naturally used for a
combined analysis of the meson electro and photoproduction
reactions.

\section{Acknowledgements}

This work was supported by the grant \textit{Russian Science
Foundation} (RSF 16-12-10267).

\end{document}